**Title:** Diversity dynamics in Nymphalidae butterflies: Effect of phylogenetic uncertainty on diversification rate shift estimates

**Authors:** Carlos Peña[1,*] and Marianne Espeland[2]
1 Laboratory of Genetics, Department of Biology, University of Turku, Turku, Finland
2 Museum of Comparative Zoology and Department of Organismic and Evolutionary Biology, Harvard University, Cambridge, USA
**Corresponding author:** mycalesis@gmail.com

**Comments:** 23 pages, 7 figures, 2 tables and 12 supplementary material files. Both authors contributed equally to this work.
**Category:** q-bio.PE

**Abstract:**
The family Nymphalidae is the largest family within the true butterflies and has been used to develop hypotheses explaining evolutionary interactions between plants and insects. Theories of insect and hostplant dynamics predict accelerated diversification in some scenarios. We investigated whether phylogenetic uncertainty affects a commonly used method (MEDUSA, modelling evolutionary diversity using stepwise AIC) for estimating shifts in diversification rates in lineages of the family Nymphalidae, by extending the method to run across a random sample of phylogenetic trees from the posterior distribution of a Bayesian run. We found that phylogenetic uncertainty greatly affects diversification rate estimates. Different trees from the posterior distribution can give diversification rates ranging from high values to almost zero for the same clade, and for some clades both significant rate increase and decrease were estimated. Only three out of 13 significant shifts found on the maximum credibility tree were consistent across more than 95% of the trees from the posterior: (i) accelerated diversification for Solanaceae feeders in the tribe Ithomiini; (ii) accelerated diversification in the genus *Charaxes*, and (iii) deceleration in the Danaina. By using the binary speciation and extinction model (BISSE), we found that a hostplant shift to Solanaceae or a codistributed character is responsible for the increase in diversification rate in Ithomiini, and the result is congruent with the diffuse cospeciation hypothesis. A shift to Apocynaceae is not responsible for the slowdown of diversification in Danaina. Our results show that taking phylogenetic uncertainty into account when estimating diversification rate shifts is of great importance, and relying on the maximum credibility tree alone potentially can give erroneous results.

# Introduction

Hostplant shifts have been invoked to be responsible for great part of the biodiversity of herbivorous insects (Mitter et al., 1988). The study of the evolution of hostplant use has spawned several theories explaining the evolutionary interactions between plants and insects (reviewed by Nyman et al, 2012). The "escape-and-radiate" hypothesis (Ehrlich & Raven, 1964) states that plants and herbivore butterflies are involved in an evolutionary arms race in which newly acquired adaptive characters in plants act as defense against herbivores. This would allow the plant lineage to escape the herbivorous pressure and diversify. Eventually, the butterflies would also acquire a character to overcome this barrier and radiate onto the available plant resource. An alternate hypothesis of herbivore diversification is the "oscillation hypothesis" (Janz, 2011) or "diffuse cospeciation" (Nyman et al., 2012) in which range expansions of plants and insects facilitate allopatric speciation and cross-colonization of insects onto related plants. This hypothesis predicts near-simultaneous diversification of plants and insects that feed on them (Nyman et al., 2012). The "resource abundance-dependent diversity dynamics" hypothesis states that plant groups that are common and widely distributed will host a higher diversity of herbivores by facilitating their speciation over wider geographic distributions. According to this hypothesis, there should be a time lag between the diversification of hostplants and insects that feed on them (Nyman et al., 2012).

The butterfly family Nymphalidae has been an important taxon for developing some of the mentioned hypotheses. Nymphalidae contains around 6000 species (van Nieukerken et al., 2011), and is the largest family within the true butterflies. The family most likely originated around 94 MYA in the mid Cretaceous. Diversification of the group began in the Late Cretaceous and most major radiations (current subfamilies) appeared shortly after the Cretaceous-Paleogene (K-Pg) boundary (Heikkilä et al. 2012). Several studies have used calibrated phylogenies and diversification models to reconstruct the evolutionary history of the group to identify patterns of accelerated or decelerated diversification of some Nymphalidae clades (Elias et al., 2009; Fordyce 2010; Wahlberg et al., 2009; Heikkilä et al., 2012). It has for example been suggested that climate change in the Oligocene and the subsequent diversification of grasses has led to diversification of the subfamily Satyrinae (Peña & Wahlberg 2008) due to the abundance of grasses over extensive geographic areas ("resource abundance-dependent diversity dynamics" hypothesis). Wahlberg et al. (2009) suggested that extinction of lineages at the K-Pg boundary was followed by increased rates of diversification in the lineages that survived this event. Fordyce (2010) found increased diversification rates in some Nymphalidae lineages after a major hostplant shift, which appears to be in agreement with Ehrlich & Raven (1964) "escape-and-radiate" model of diversification.

Although it has been suggested that part of the great diversity of Nymphalidae butterflies is a result of hostplant-insect dynamics, it is necessary to use modern techniques to investigate whether the diversification patterns of Nymphalidae are in agreement with the theoretical predictions. It is necessary to test whether the overall diversification pattern of Nymphalidae is congruent with events of sudden diversification bursts due to hostplant shift ("radiate-and-escape" hypothesis, eg. Fordyce 2010), climatic events and shifts to closely related hostplants ("diffuse cospeciation hypothesis", eg. McLeish et al. 2007; Voje et al., 2009) or resource-abundance dynamics hypothesis (eg. Peña & Wahlberg, 2008). If patterns of diversification bursts are identified, it is necessary to test whether this is caused by a shift onto chemically different hostplant groups, climatic change, geography, ecological opportunity, or other factors.

In this study, we used a time calibrated genus-level phylogenetic hypothesis for Nymphalidae butterflies (taken from Wahlberg, et al. 2009) to investigate patterns of diversification. We applied MEDUSA (modelling evolutionary diversity using stepwise AIC, Alfaro et al., 2009; Harmon et al. 2011), a recently developed statistical method, to study the diversification pattern of Nymphalidae butterflies. Particularly we wanted to study the effects of phylogenetic uncertainty and modified the current MEDUSA method to take this into account (Multi-MEDUSA). We also tested whether hostplant association dynamics can explain the diversification patterns of component Nymphalidae lineages by testing whether character states of hostplant use affected the diversification pattern of those lineages employing the method BiSSE as implemented in the R package *diversitree* (FitzJohn, 2012).

# Material and methods:

### Data

For analyses, we used the phylogenetic trees from the study of Wahlberg, et al. (2009) that were generated using DNA sequence data from 10 gene regions for 398 of the 540 valid genera in Nymphalidae. We employed Wahlberg, et al. (2009) maximum clade credibility tree (*supp. mat. 01*) as well as a random sample of 1000 trees from their BEAST run after burnin. Their original BEAST run was for 40 million generations. We used a burnin of 25 million generations and took a random sample of 1000 trees using Burntrees v.0.1.9 (http://www.abc.se/~nylander/) (*supp. mat. 02*) in order to correct for phylogenetic uncertainty when performing the diversification analyses.

We compiled species richness data for Nymphalidae genera from several sources including the specialist-curated lists on http://tolweb.org, Lamas (2004) and curated lists of Global Butterfly Names project http://www.ucl.ac.uk/taxome/gbn/. We assigned the species numbers of genera not included in the phylogeny to the closest related genus that was included in Wahlberg et al. (2009) study (*supp. mat. 03*).

Hostplant data for Nymphalidae species were compiled from several sources including Ackery (1988), HOSTS database (http://bit.ly/YI7nwW), Dyer & Gentry (2002) and others (*supp. mat. 04; supp. mat. 05*) for a total of 6586 hostplant records, including 428 Nymphalidae genera and 143 plant families and 1070 plant genera. It was not possible to find hostplant data for 35 butterfly genera (*supp. mat. 04*).

### Analyses of Diversification

We used the statistical software R version 2.15.1 (R Core Team, 2012) in combination with the APE (Paradis et al., 2004), GEIGER (Harmon et al., 2008) and *diversitree* (FitzJohn, 2012) packages along with our own scripts to perform the analyses (included as supplementary materials). All analyses were run on the 1000 random trees from Wahlberg et al. (2009) as well as on the maximum credibility tree.

**LTT plots**

We obtained semilogarithmic lineages-through-time (LTT) plots after removing outgroups and including the maximum credibility tree and 95% credibility interval obtained by random sampling of 1000 trees from the posterior distribution of the BEAST run after burnin from Wahlberg et al. (2009).

**MEDUSA**

We analyzed patterns of diversification in Nymphalidae by using Turbo-MEDUSA version 1.0 (Harmon et al., 2011) on the maximum credibility tree from Wahlberg et al. (2009). Turbo-MEDUSA, and the original algorithm MEDUSA, fit alternative birth-death likelihood models to a phylogenetic tree in order to estimate changes in net diversification rates along branches. MEDUSA estimates likelihood and AIC scores for the simplest birth-death model, with two parameters ($b$: speciation and $d$: extinction). The AIC scores of the two-parameter model are then compared with incrementally more complex models until the addition of parameters do not improve the AIC scores beyond a cutoff value. MEDUSA finds the likelihood of the models after taking into account branch lengths and number of species per lineage (Alfaro et al., 2009). To our knowledge MEDUSA and Turbo-MEDUSA have so far only been run on a single tree, usually the maximum credibility tree, which makes the assumption that this tree is correct. We wanted to study the effects of phylogenetic uncertainty on estimation of diversification rate shifts and therefore used custom-made scripts to run Turbo-MEDUSA across 1000 random genus-level trees from the posterior distribution (Multi-MEDUSA, *supp. mat. 06*) and summarize the estimated changes in diversification rates for nodes across all trees. Patterns of change in diversification rates are significant if they are found at the same node in at least 95% of the trees. We also expected to find similar $r$ and *epsilon* values across the 1000 trees for the nodes where changes in diversification tempo occurs. We let MEDUSA estimate up to 25 turnover points in our trees.

**BiSSE**

We tested whether diversification of Nymphalidae lineages is driven by hostplant by using the "binary state speciation and extinction" (BiSSE; Maddison, Midford, Otto, 2007) bayesian approach as implemented in the R package *diversitree* (FitzJohn, 2012). MuSSE (FitzJohn, 2012) is designed to examine the joint effects of two or more traits on speciation. Because most of Nymphalidae butterflies are restricted to use one plant family as hostplant, the character states can be coded as presence/absence, for which the BiSSE analysis is better suited. BiSSE was designed to test whether a binary character state has had any effect on increased diversification rate for a clade (Maddison et al., 2007). We used our compiled data of hostplant use to produce binary datasets for the characters "feeding on the plant family Solanaceae or Apocynaceae" (*supp. mat. 07*) which are the main hostplants of the diverse Ithomiini butterflies and closest relatives (Willmott & Freitas 2006) and Danaini (Apocynaceae). We analyzed the data using BiSSE employing markov chain monte carlo algorithm taking into account missing taxa by using the parameter "sampling factor" (*sampling.f*) in *diversitree*. We also used constrained analyses forcing no effect of hostplant use on diversification and likelihood ratio tests to find out whether the hypothesis of effect on diversification has a significantly better likelihood than the null hypothesis (no effect).

# Results

**LTT plot**

LTT plots of the 1000 randomly chosen trees, along with the maximum credibility tree, from the posterior distribution of trees from Wahlberg et al. (2009) are shown in Figure 1. This figure shows that the accumulation of lineages in Nymphalidae is congruent with a pattern of increase in diversification rate starting at around 50 Mya. However, it should be noted that we used a genus level phylogeny with very incomplete sampling at the species level. Thus, the slowdown of diversification that is apparent during the last 20 MYA is an artifact and should be interpreted very carefully.

**MEDUSA**

The MEDUSA analysis on the MCT tree in combination with richness data estimated 13 changes in the tempo of diversification in Nymphalidae history (Figure 2; Table 1). The estimated cutoff value (corrected threshold) of AICc scores for selecting the optimal model was estimated as 7.8 units. The background diversification rate for Nymphalidae was estimated as $r = 0.081$ lineages per Million of years. Some of the 13 changes in diversification correspond to rate increases in very species-rich genera: *Ypthima* ($r = 0.264$), *Charaxes* ($r = 0.251$), *Callicore* + *Diaethria* ($r = 0.220$), *Pedaliodes* ($r = 0.196$) and *Taenaris* ($r = 0.238$). We found rate increases for other clades as well such as: Lethina + Mycalesina ($r = 0.130$), Oleriina + Ithomiina + Napeogenina + Dircennina + Godyrina ($r = 0.181$), Euptychiina + Pronophilina + Maniolina + Satyrina ($r = 0.114$), Phyciodina in part ($r = 0.227$) and Satyrina ($r = 0.220$).

**Phylogenetic uncertainty in the Multi-MEDUSA approach**

We found that the analyses by MEDUSA on the 1000 trees did not estimate the same diversification shifts as in the MCT (all shifts found by MEDUSA on the 1000 trees are provided in *supp. mat. 08*). In order to obtain the diversification shifts that were estimated in most of the 1000 trees, we plotted the diversification shifts (index number) versus number of trees containing that particular diversification shift (Figure 3, *supp. mat. 09-10*) as estimated by MEDUSA. Besides the root, there were three diversification shifts found in more than 95% of the trees: (*i*) diversification rate increase in the genus *Charaxes*; (*ii*) rate increase in Ithomiini subtribes Oleriina + Ithomiina + Napeogenina + Dircennina + Godyrina, while (*iii*) slowed diversification in part of Danaini (Figure 3).

We obtained mean and standard deviation statistics for the diversification values found on the shifts on the 1000 trees (*supp. mat. 09*). We found that some of the changes in diversification rate values had great variation across the posterior distribution of trees. A boxplot of the diversification rate values estimated for the clades that appear in the MCT shows that some shifts are estimated as increased or slowed diversification pace depending on the tree used for analysis (Figure 4). This variation is especially wide for the clade formed by the genera *Magneuptychia* and *Caeruleuptychia* because MEDUSA estimated diversification values from three times the background diversification rate ($r = 0.2755$) to almost zero ($r = 2.4e-07$). The diversification rates estimates for the root (background diversification rate) and the clades (Tirumala + Danaus

+ Amauris + Parantica + Ideopsis + Euploea + Idea) and (Oleriina + Ithomiina + Napeogenina + Dircennina + Godyrina) are relatively consistent across the 1000 trees (Figure 4). It is also evident that not all the diversification shifts estimated on the MCT are consistently recovered in most of the 1000 trees. Some of the splits in the MCT are recovered in very few trees, for example the split for the clade (Euptychiina + Pronophilina + Satyrina + Maniolina) (*see supp. mat. 09-10*).

**BiSSE**

The MEDUSA analyses taking into account phylogenetic uncertainty estimated a diversification rate increase in part of the clade Ithomiini across more than 95% of the trees. Our BiSSE analysis found a positive effect of the character state "feeding on Solanaceae" on the diversification rate on part of Ithomiini (Oleriina + Ithomiina + Napeogenina + Dircennina + Godyrina) (Figure 5). The markov chain monte carlo algorithm was run for 10000 generations discarding the first 7500 as burnin. The estimated mean diversification rate for taxa that do not feed on Solanaceae was $r = 0.10$ while the diversification rate for the Solanaceae feeders was $r = 0.14$ (see Figure 6 for a boxplot of speciation and extinction values for the 95% credibility intervals). We constrained the BiSSE likelihood model to force equal rates of speciation for both character states in order to test whether the model of different speciation rates is a significantly better explanation for the data. A likelihood ratio test found that the model for increased diversification rate for nymphalids feeding on Solanaceae is a significantly better explanation than this character state having no effect on diversification ($p < 0.001$) (Table 2; character states available in *supp. mat. 07*, code in *supp. mat. 11*, and mcmc run in *supp. mat. 12*). A BiSSE analysis to test whether the trait "feeding on Apocynaceae" had any effect on increased diversification rates found similar speciation rates for lineages feeding on Apocynaceae and other plants (Figure 7). It has been shown that BiSSE performs poorly when certain conditions are met (Davis et al., 2013). However, our data has adequate number of taxa under analysis (more than 300 tips), adequate speciation bias (between 1.5x and 2.0x), character state bias (around 8x) and extinction bias (around 4x) for the analysis of Solanaceae hostplants. Thus, BiSSE is expected to produce robust results (Davis et al., 2013).

# Discussion

**Effects of phylogenetic uncertainty on the performance of MEDUSA**

The MEDUSA method has been used to infer changes in diversification rates along a phylogenetic tree. Since its publication (Alfaro et al., 2009), the results of using MEDUSA on a single tree, the maximum clade credibility tree, have been used for generation of hypotheses and discussion (e.g. Litman et al., 2011; Heikkilä et al., 2012; Ryberg & Matheny, 2012). However, we found that MEDUSA estimated different diversification shifts and different rates of diversification when phylogenetic uncertainty was taken into account by using MEDUSA on a random sample of trees from the posterior distribution of a Bayesian run. We found that some diversification splits, estimated on the Nymphalidae maximum probability tree, were found in a very small percentage of the 1000 randomly sampled trees from the posterior distribution (Figure 3). We also found that, even though MEDUSA could estimate the same diversification splits on two or more trees, the estimated diversification rates could vary widely (Figure 4). For example, in our Nymphalidae trees, we found that the split for *Magneuptychia* and *Caeruleuptychia* had a variation from *r* = 0.2755, higher than the background diversification rate, to almost zero. This means that observed patterns and conclusions can be completely contradictory depending on tree choice.

In this study, the effect of phylogenetic uncertainty on the inferred diversification splits by MEDUSA is amplified because some Nymphalidae taxa appear to be strongly affected by long-branch attraction artifacts (Peña et al., 2011). Thus, the Bayesian runs are expected to recover alternative topologies on the posterior distribution of trees, resulting in low support and posterior probability values for the nodes. For example, posterior probability values for clades in Satyrini are very low (0.5 to 0.6; Wahlberg et al., 2009). As a result, MEDUSA inferred a diversification rate increase for part of Satyrini in the maximum credibility tree, but this was recovered only in 17% of the trees from the posterior distribution.

If there is strong phylogenetic signal for increases or decreases in diversification rates for a node, it is expected that these splits would be inferred by MEDUSA in most of the posterior distribution of trees. However, weak phylogenetic signal for some nodes can cause some clades to be absent in some trees and MEDUSA will be unable to estimate any diversification shift (due to a non-existent node). This is the reason why MEDUSA estimated diversification rate splits in more than 95% of the posterior distribution of trees for only three splits: the genus *Charaxes*, part of Danaini and part of Ithomiini (Figure 3), while estimating splits for other lineages in only a fraction of the posterior distribution of trees.

The clade Ithomiini and the non-basal Danaids are well supported by high posterior probability values in Wahlberg et al. (2009; supplementary information, fig. 3S). Therefore our MEDUSA analyses recovered an increase in diversification rate in more than 95% of the posterior distribution of trees (Figure 3).

**Hostplant use and diversification in Nymphalidae**
### Ithomiini
Keith Brown suggested that feeding on Solanacaeae was an important event in the diversification of Ithomiini butterflies (Brown 1987). The Ithomiini butterflies are exclusively Neotropical and most species feed on Solanaceae hostplants during larval stage (*supp. mat. 04*; Willmott & Freitas, 2006). Optimizations of the evolution of hostplant use on phylogenies evidence a probable shift from Apocynaceae to Solanaceae (Brower et al., 2006; Willmott & Freitas, 2006). Fordyce (2010) found that the Gamma statistics, a LTT plot of an Ithomiini phylogeny and the fit of the density-dependent model of diversification are consistent with a burst of diversification in Ithomiini following the shift from Apocynaeae to Solanacaeae.

In this study, we investigated whether the strong signal for an increase in diversification rate for Ithomiini (found by MEDUSA) can be explained due to the use of Solanaceae plants as hosts during larval stage. For this, we used a Bayesian approach (BiSSE; FitzJohn et al., 2009) to test whether the trait "feeding on Solanaceae" had any effect on the diversification of the group.

Our BiSSE analysis, extended to take into account missing taxa, shows a significantly higher net diversification rate for Ithomiini taxa, which can be attributed to the trait "feeding on Solanaceae hostplants" (Figure 5). This is in agreement with the findings of Fordyce (2010) using other statistical methods. Due to the fact that Ithomiini are virtually the only nymphalids using Solancaeae as hostplants (except for *Hypanartia*, *Vanessa* and *Acraea*; *supp. mat.04*), it is possible that the trait responsible for a higher diversification of Ithomiini might not be the hostplant character. As noted by Maddison et al. (2007), the responsible trait might be a codistributed character such as a trait related to the ability to digest secondary metabolites.

Solanaceae plants contain chemical compounds and it has been suggested that the high diversity of Ithomiini is consistent with the "escape-and-radiate scenario" due to a shift onto Solanaceae (Fordyce 2010) and radiation scenarios among chemically different lineages of Solanaceae plants (Brown, 1987; Willmott & Freitas, 2006). According to this theory, the shift from Apocynaceae to Solanaceae allowed the Ithomiini to invade newly available resources due to a possible key innovation that allowed them cope with secondary metabolites of the new hosts. Although some herbivores might shift to inferior hostplants under certain conditions (e.g. predation), this is expected to be unlikely in herbivores adapted to feed on chemically protected plant species by secondary compounds (De-Silva et al., 2011), such as Solanaceae species. Additional studies are needed to identify the actual enzymes that Ithomiini species might be using for detoxification of ingested food. A detoxification mechanism has been found in Pieridae larvae feeding on Brassicales hosts. The nitrile-specifier protein is considered a key innovation for high diversification in Pieridae butterflies (Wheat et al., 2007).

The diffuse cospeciation hypothesis predict almost identical ages of insects and their hostplants, while the "resource abundance-dependent diversity" and the "escape-and-radiate" hypotheses state that insects diversify after their hostplants (Ehrlich & Raven, 1964; Janz, 2011; Nyman et al. 2012). Wheat et al. (2007) found strong evidence for a model of speciation congruent with Ehrlich and Raven's hypothesis in Pieridae butterflies due to, in addition to the identification of a key innovation, a burst of diversification in glucosinolate-feeding taxa shortly afterwards (with a lag of ~10 My). According to a recent dated phylogeny of the Angiosperms (Bell et al. 2010), the family Solanaceae split from its sister group about 59 (49-68) MYA and diversification started (crown group age) around 37 (29-47) MYA. Wahlberg et al. (2009) give the corresponding ages for Ithomiini as 45 (39-53) and 37 (32-43) MYA, respectively. Thus, it is evident that the Solanaceae and Ithomiini diversified around the same time, during the Late Eocene and Oligocene, and this is congruent with the diffuse cospeciation hypothesis.

The uplift of the Andes was a major tectonic event that underwent higher activity during the Oligocene (Somoza, 1998). This caused climatic changes in the region that affected the flora and fauna of the time, which coincides with the diversification of modern montane plant and animal taxa (Hoorn et al., 2010) including Ithomiini butterflies and Solanaceae hostplants. Moreover, all Solanaceae clades currently present in New World originated in South America (Olmstead, 2013) as well as Ithomiini butterflies (Wahlberg et al., 2009). Therefore, there is evidence for a process of "diffuse cospeciation" of Ithomiinae and hostplants.

**Danaini**

Our Multi-MEDUSA approach gave a significant slowdown in diversification rate in the subtribe Danaina of the Danini. Both Danaina and the sister clade Euploeina feed mainly on Apocynaceae and thus a hostplant shift should not be responsible for the observed slowdown of diversification in the Danaina. As expected, our BiSSE analysis of Apocynaceae feeders shows that there is no effect of feeding on this plant family on the diversification rates of Nymphalidae lineages. Many of the Danaina are large, strong fliers, highly migratory and involved in mimicry rings. Among them is for example the monarch (*Danaus plexippus*), probably the most well known of all migratory butterflies. The causes for a lower diversification rate in the Danaina remains to be investigated, but their great dispersal power might be involved in preventing allopatric speciation. It has been found in highly vagile species in the nymphalid genus *Vanessa* that dispersal has homogenized populations due to gene flow, as old and vagile species seem to be genetically homogeneous while younger widespread species show higher genetic differentiation in their populations (Wahlberg & Rubinoff, 2011).

*Charaxes*

The genus *Charaxes* contain 193 species distributed in the Old World with highest diversity in the Afrotropical region. These butterflies are also very strong fliers, but contrary to Danaina, which are specialized Apocynaceae feeders, *Charaxes* are known to feed on at least 28 plant families in 18 orders (Ackery, 1988) and some species appear to be polyphagous (Müller et al., 2010). Aduse-Poku et al. (2009) showed that most of the diversification of the genus occurred from the late Oligocene to the Miocene when there were drastic global climatic fluctuations, indicating that the diversity mainly is driven by climate change. Müller et al. (2010) found that climatic changes during the Pliocene, Pleistocene as well as dispersal and vicariance might have been responsible for the high diversification of the genus.

**Satyrini**

The diverse tribe Satyrini radiated simultaneously with the radiation of their main hostplant, grasses, during the climatic cooling in the Oligocene (Peña & Wahlberg, 2008). Thus, it is somewhat surprising that part of Satyrini (the subtribes Euptychiina, Satyrina and Pronophilina) were found to have accelerated diversification in only 17% of the trees from the posterior distribution. Although this can be attributed to low phylogenetic signal (posterior probability value = 0.6 for this clade in Wahlberg et al., 2009), the clade Satyrini is very robust (posterior probability value = 1.0 for this clade in Wahlberg et al., 2009) and MEDUSA failed to identify any significant accelerated diversification rate for Satyrini. It appears that the radiation of Satyrini was not remarkable quick to be picked up by MEDUSA.

The origin of the tribe Satyrini is not completely clear (originated either in the Neotropical or Eastern Palaearctic, Oriental and/or Indo-Australian regions (Peña et al., 2011) and their radiation involved colonizing almost all continents starting from their place of origin. For Satyrini butterflies to colonize new areas, it was necessary that their hosplants had preceded them. It takes time to colonize a new continent, so we can speculate that it took even longer for lineages in Satyrini to colonize new areas and reach their current distributions. Then, it should be expected that the diversification of Satyrinae occurred in a stepwise manner, with pulses or bursts of diversification for certain lineages but unlikely for the tribe Satyrini as a whole.

**Conclusions**

We found that even though MEDUSA estimated several diversification shifts in the maximum credibility tree of Nymphalidae, only a few of these splits were found in more than 95% of the trees from the posterior distribution. In the literature, it is common practice that conclusions are based on the splits estimated on the maximum credibility tree. However, by using a Multi-MEDUSA approach, we found that some of this splits might be greatly affected by phylogenetic uncertainty. Moreover, some of these splits can be recovered either as increases or decreases in diversification rate depending on the tree from the posterior distribution that was taken for analysis. This means that contradictory conclusions would be made if only the maximum credibility tree was used for analysis.

Our Multi-MEDUSA approach to perform analyses on the posterior distribution of trees found strong support for an increase in diversification rate in the tribe Ithomiini and the genus *Charaxes*, and for a decrease in diversification rate in the subtribe Danaina. Due to phylogenetic uncertainty, we did not obtain strong support for other diversification splits in Nymphalidae. Our BiSSE analysis found that the trait "feeding on Solanaceae", or a codistributed character, is responsible for the higher diversification rate of the Ithomiini, but the trait "feeding on Apocynaceae" is not responsible for the slowdown of diversification in the Danaina . Ithomiini and Solanaceae diversified near simultaneously, which is in agreement with the diffuse co-speciation hypothesis (Janz 2011, Nyman et al. 2012).

## Acknowledgments
We are thankful to Mark Cornwall for help with the script to extend MEDUSA to include phylogenetic uncertainty, Niklas Wahlberg for commenting on the manuscript and giving us the posterior distribution of trees, Jessica Slove Davidson and Niklas Janz for access to their hostplant data. The study was supported by a Kone Foundation grant (awarded to Niklas Wahlberg), Finland (C. Peña) and the Research Council of Norway (grant no. 204308 to M. Espeland).

**Figure legends**

**Figure 1.** Lineage through time plot for the maximum credibility tree (black) and 1000 random trees from the posterior distribution (coloured lines) of the Nymphalid phylogeny. Since this is a genus level phylogeny the observed slowdown of diversification the last 20 MYA is an artefact and should be disregarded.

**Figure 2.** Results of the MEDUSA analysis run on the maximum credibility tree. Rate shifts were estimated for the following nodes: 1) Background rate, 2) Limenitidinae + Heliconiinae, 3) *Ypthima*, 4) *Charaxes*, 5) Mycalesina + Lethina, 6) Ithomiini in part, 7) Satyrini in part, 8) Phyciodina in part, 9) Danaini in part, 10) *Caeruleuptychia* + *Magneuptychia*, 11) Satyrina, 12) *Callicore* + *Diaethria*, 13) *Pedaliodes*, 14) *Taenaris*.

**FIgure 3.** Results of the Multi-MEDUSA analysis on 1000 random trees from the posterior distribution of the Nymphalidae phylogeny. Bars show the number of trees where MEDUSA found significant increases or decreases in diversification rates for tips or clades in Nymphalidae. Tip and clade indexes are detailed in supplementary material 10.

**Figure 4.** Boxplot of the range of diversification values for tips or clades estimated by MEDUSA on the 1000 random trees from the posterior distribution of the Nymphalidae phylogeny. Those tips or clades present on the maximum credibility tree are shown.

**Figure 5.** BiSSE analysis of diversification of nymphalids due to feeding on Solanaceae hostplants. Speciation and diversification rates are significantly higher in Solanaceae feeders.

**Figure 6.** Boxplot of speciation (lambda), extinction (mu) and transition (q) parameter values of the BiSSE analysis on diversification due to feeding on Solanaceae hostplants.

**Figure 7.** BiSSE analysis of diversification of nymphalids due to feeding on Apocynaceae hostplants. There is no effect of this trait either on speciation or diversification rates.

**Table 1.** Significant diversification rate shifts found in the Turbo-MEDUSA analysis of the Nymphalid maximum credibility tree. Split.Node = node number, r = net diversification rate, epsilon = relative extinction rate, LnLik.part = log likelihood value

**Table 2.** Likelihood ratio test between the model of increased diversification of nymphalids feeding on Solanaceae against a model forcing equal speciation rates (no effect on diversification).

# Figures

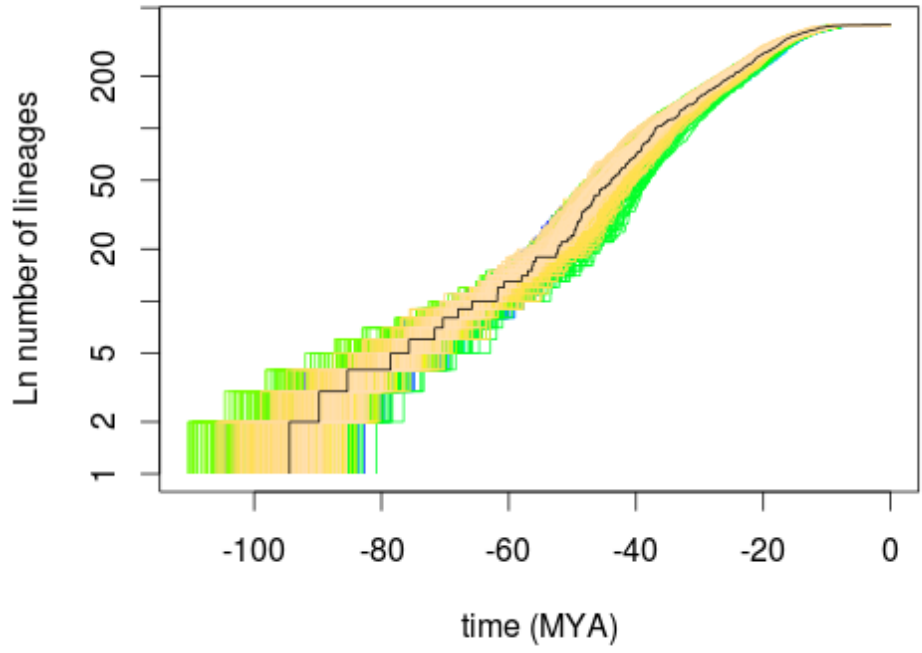

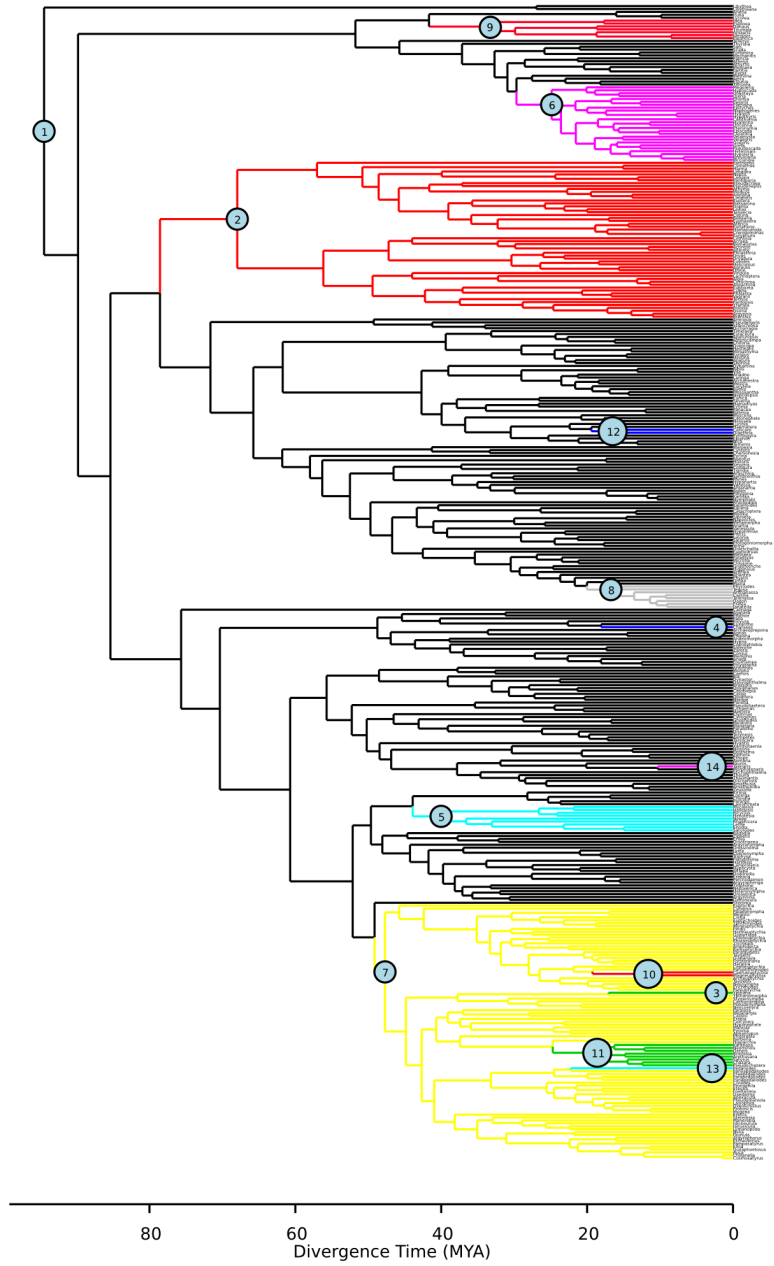

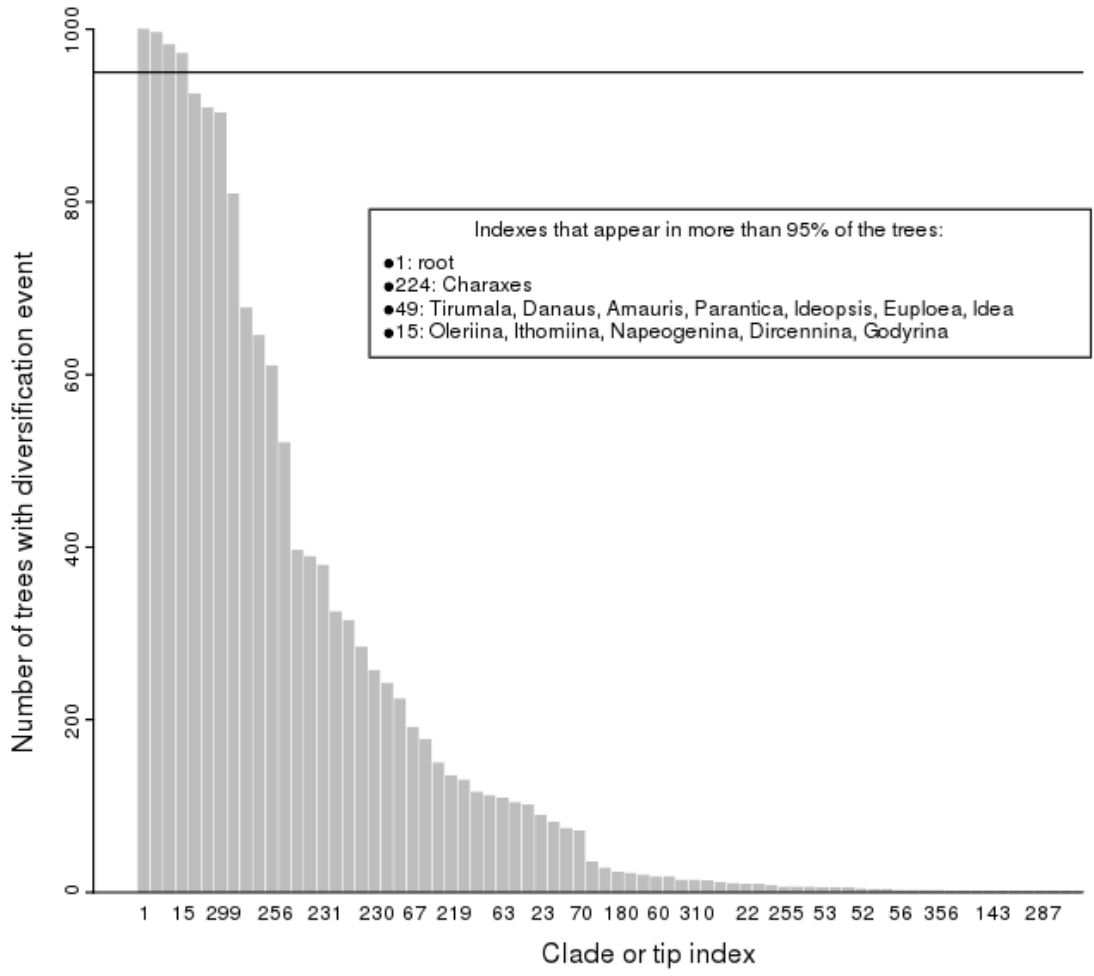

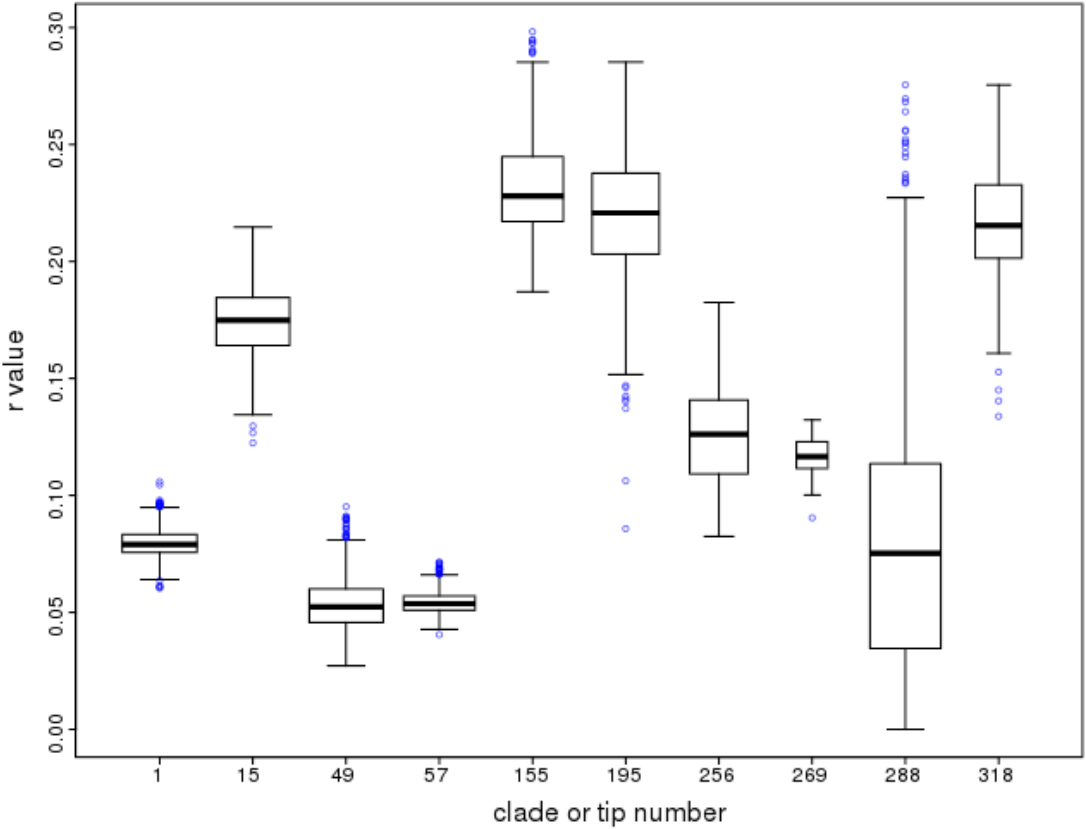

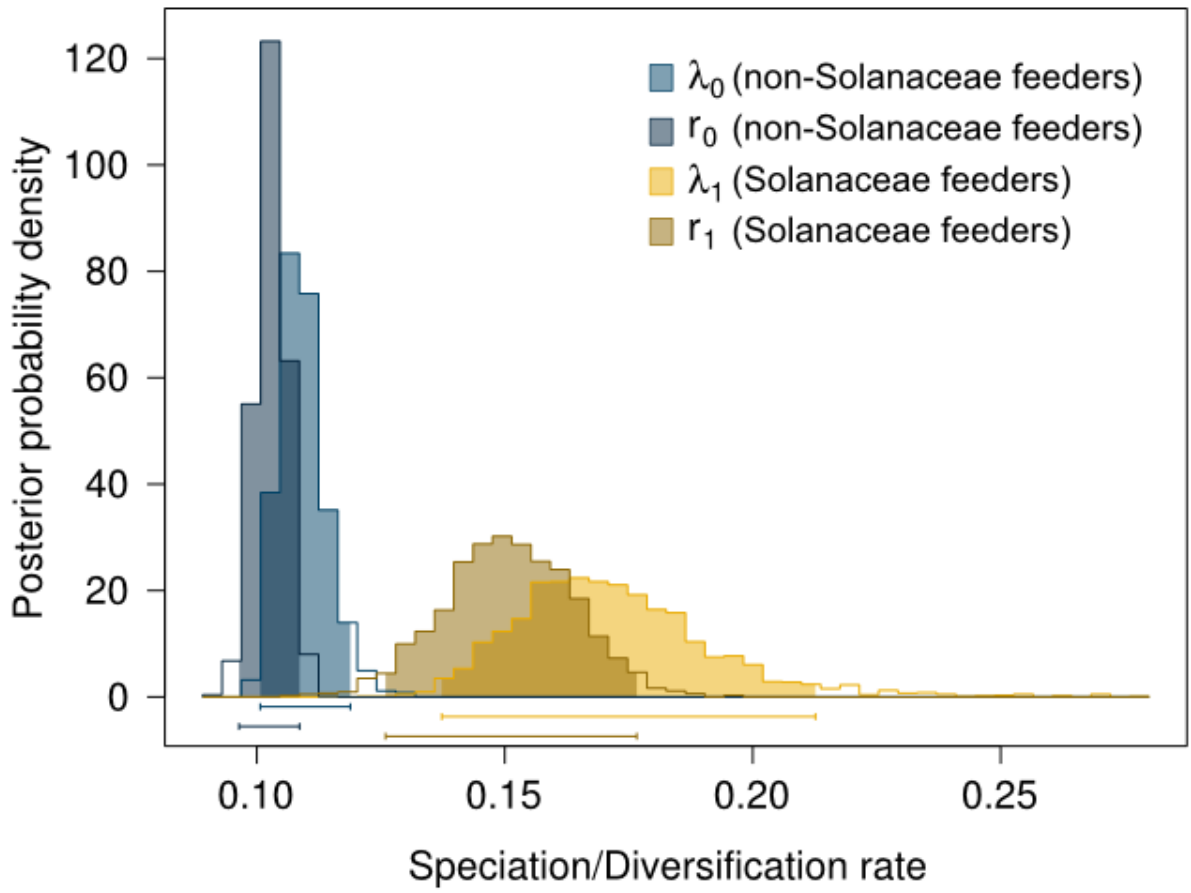

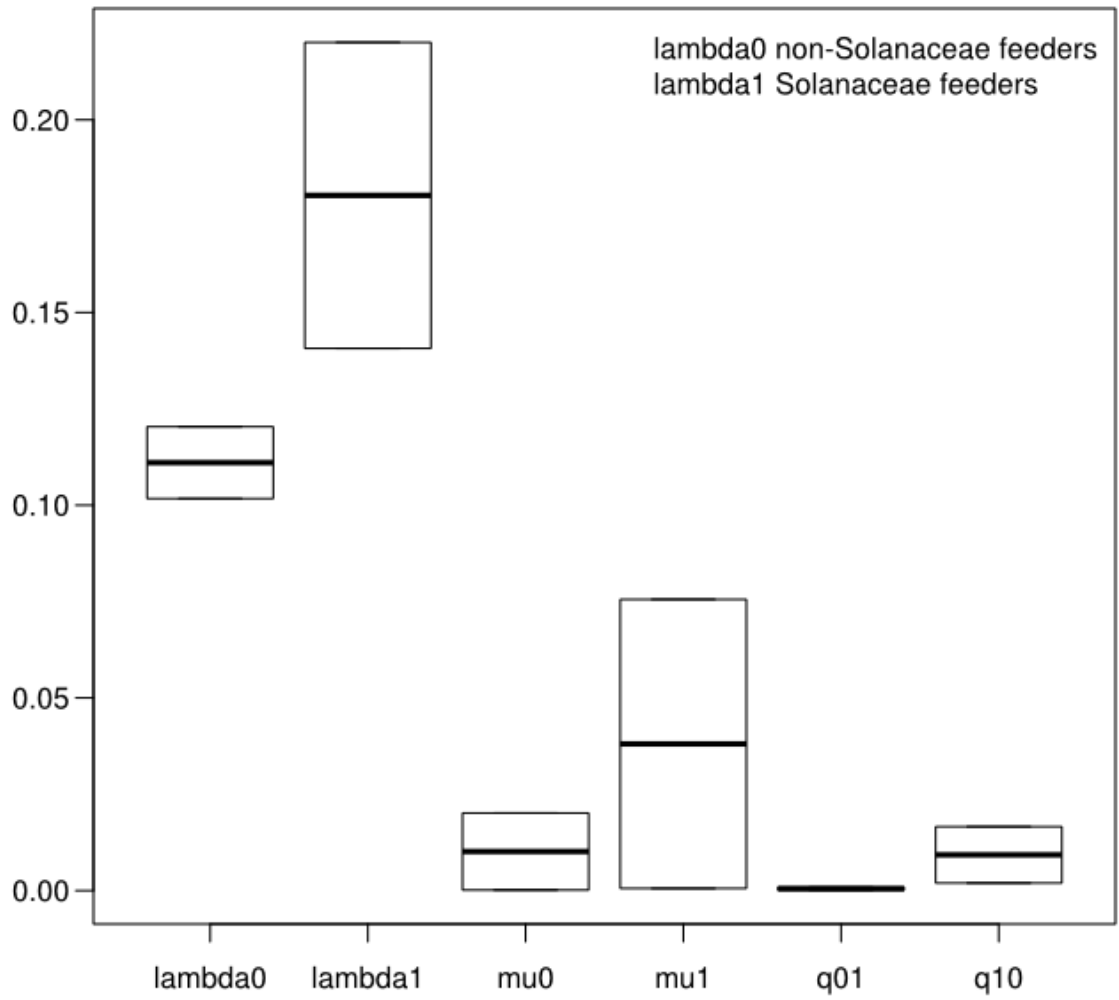

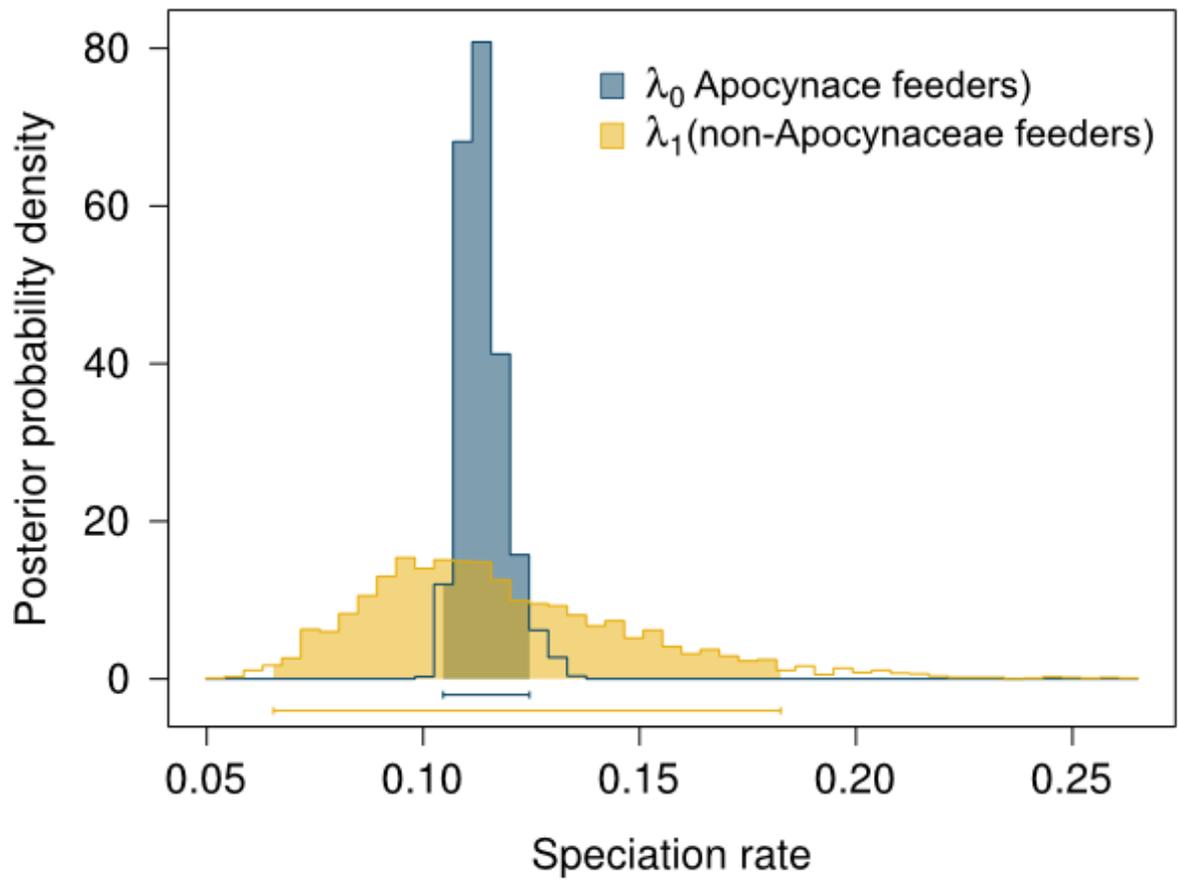

# Tables

Table 1
Diversity analysis of Turbo-MEDUSA on the Nymphalidae MCT.

|    | Split.node | r   | epsilon   | LnLik.part | Taxa |
|----|------------|-----|-----------|------------|------|
| 1  | 399 | 0.0811 | 0.22222 | -1262.0629 | Nymphalidae |
| 2  | 455 | 0.0539 | 0.94482 | -417.5449 | Limenitidinae + Heliconiinae |
| 3  | 299 | 0.2642 | 0.55441 | -6.3058 | *Ypthima* |
| 4  | 224 | 0.2505 | 0.52112 | -6.2601 | *Charaxes* |
| 5  | 654 | 0.1297 | 0.62404 | -67.778 | Lethina + Mycalesina |
| 6  | 413 | 0.1812 | 0.086651 | -147.4009 | Oleriina + Ithomiina + Napeoge |
| 7  | 667 | 0.1143 | 4.53E-08 | -459.7479 | Euptychiina + Pronophilina + N |
| 8  | 553 | 0.2270 | 2.80E-07 | -41.4222 | Phyciodina in part |
| 9  | 447 | 0.0555 | 0.93003 | -47.0677 | Danaini in part |
| 10 | 686 | 0.0685 | 0.96681 | -12.1266 | *Caeruleuptychia* + *Magneupty* |
| 11 | 716 | 0.2196 | 2.97E-06 | -46.6369 | Satyrina |
| 12 | 593 | 0.2196 | 7.68E-06 | -11.2686 | *Callicore* + *Diaethria* |
| 13 | 355 | 0.1964 | 0.56972 | -6.2013 | *Pedaliodes* |
| 14 | 377 | 0.2377 | 0.55632 | -4.1986 | *Taenaris* |

Table 2. Likelihood ratio test

|  | Df | lnLik | AIC | ChiSq | P |
|---|---|---|---|---|---|
| full | 6 | -1612.3 | 3236.6 | | |
| equal.lambda | 5 | -1619.4 | 3248.8 | 14.222 | 0.00016 |

# Supplementary Material (R code, phylogenetic trees, hostplant data, etc.)

Available at http://dx.doi.org/10.6084/m9.figshare.639208